# Deep Representation Learning for Unsupervised Clustering of Myocardial Fiber Trajectories in Cardiac Diffusion Tensor Imaging

Mohini Anand[1], Xavier Tricoche[1]

[1] Purdue University, West Lafayette, IN, USA
{anand130, xmt}@purdue.edu

**Abstract.** Understanding the complex myocardial architecture is critical for diagnosing and treating heart disease. However, existing methods often struggle to accurately capture this intricate structure from Diffusion Tensor Imaging (DTI) data, particularly due to the lack of ground truth labels and the ambiguous, intertwined nature of fiber trajectories. We present a novel deep learning framework for unsupervised clustering of myocardial fibers, providing a data-driven approach to identifying distinct fiber bundles. We uniquely combine a Bidirectional Long Short-Term Memory network to capture local sequential information along fibers, with a Transformer autoencoder to learn global shape features, with pointwise incorporation of essential anatomical context. Clustering these representations using a density-based algorithm identifies 33 to 62 robust clusters, successfully capturing the subtle distinctions in fiber trajectories with varying levels of granularity. Our framework offers a new, flexible, and quantitative way to analyze myocardial structure, achieving a level of delineation that, to our knowledge, has not been previously achieved, with potential applications in improving surgical planning, characterizing disease-related remodeling, and ultimately, advancing personalized cardiac care.

**Keywords:** Cardiac DTI, Deep Learning, Unsupervised Clustering

## 1 Background

The heart's fine-grained fiber organization underpins both normal function and disease, yet many macroscopic models overlook the detailed arrangement of cardiomyocytes. Given the pivotal role of cardiomyocyte orientation in both normal physiology and disease progression [25], our novel framework bridges the gap between mesoscopic fiber orientations and their organ-level implications by integrating myocardial fiber architecture features, extracted from diffusion tensor imaging (DTI), to achieve a more comprehensive and spatially resolved understanding of whole-heart structure.

DTI maps water diffusion in tissues to reveal myocyte orientations and reconstructs 3D fiber pathways to create comprehensive maps of cardiac architecture through fiber tractography [5]. It has seen diverse applications—including validation of DTI's ability to reproduce myocardial architecture [8], refinement of acquisition protocols [12], studies of fetal cardiac development [11], assessment of post-intervention remodeling [22], study of infarcted hearts [15], and the development of novel metrics for quantifying scar tissue in 3D space [13].

However, current cardiac DTI studies typically focus on localized regions and visual comparisons, largely because whole-heart DTI data is extremely complex, with overlapping trajectories and ambiguous endpoints. Consequently, these approaches neglect the global, interconnected organization of myocardial fiber groups, limiting insights into how regional microstructural changes affect organ-level function. Yet, analyzing these trajectories alongside orientation and tissue properties could reveal essential patterns. We address this gap by positing that a well-designed whole-heart streamline clustering approach can unlock a more comprehensive structural analysis of the myocardium.

Neuroimaging has successfully used DTI tractography clustering to study white matter architecture, grouping similar tracts through shape simplification [7], feature extraction (e.g., curvature, torsion, and landmark proximity) [14], or image-like representations [27]. Unsupervised deep learning discovers patterns without explicit labels by comparing fiber trajectories [29], reconstructing fiber pathways [28], or filtering out implausible fibers [9]. More recent transformer-based methods [24] automatically learn embeddings for fiber classification, frequently incorporating anatomical knowledge. These successes in brain DTI suggest that similar strategies could be adapted to unravel the complex fiber populations in the whole heart.

Despite the potential of DTI tractography, whole-heart fiber clustering remains largely unexplored in cardiology. Existing approaches often focus on slice-based analyses [26], omit crucial orientation information like the Helical Angle [6], analyze only localized regions [30], or rely on manual refinement [15, 21]. These methods struggle with the heart's complex 3D architecture, where fibers extensively overlap without clear endpoints. Unlike in the brain—where distinct branching aids tract separation—cardiac fibers lack strong anatomical landmarks, and ground-truth labels remain unavailable. These challenges motivate a global, data-driven framework that integrates geometric, and orientation features for comprehensive whole-heart fiber delineation.

We propose a novel deep-learning framework (**Fig. 1**) for whole-heart fiber trajectory clustering, robust to noise and anatomical variations. Our contributions are: 1) A unique cardiac DTI data representation incorporating per-point helical angle, transmural depth, and spatial coordinates, capturing myocyte orientation and relative spatial location; 2) To capture the complex cardiac fiber dynamics, we devise a novel pretext task of predicting future fiber points and their respective features, forcing a Bidirectional Long Short-Term Memory (BLSTM) model, well-suited for capturing sequential dependencies, to learn these properties; 3) Fusion of these sequential embeddings with global features from a Transformer autoencoder, results in a comprehensive representation; 4) Finally, we employ Hierarchical Density-Based Spatial Clustering of Applications with Noise (HDBSCAN) for clustering and outlier detection, with the potential to identify diseased regions with atypical fiber organization. Our work significantly advances whole-heart DTI analysis with a level of delineation, that to our knowledge, has not been previously achieved.

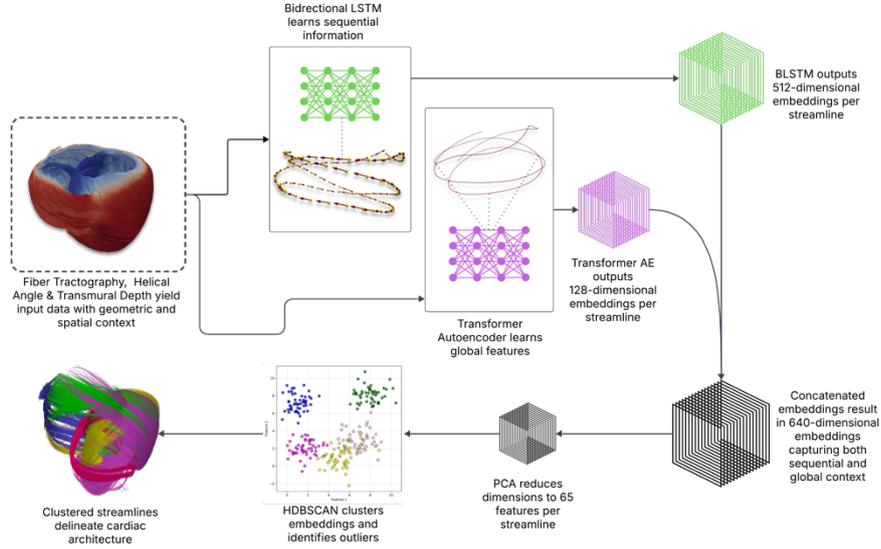

**Fig. 1.** Overview of our framework

## 2      Materials and Methods

### 2.1    Data and Feature Extraction

We utilized a publicly available atlas [16], averaging high-resolution DTI data from nine ex vivo canine hearts, acquired with a 1.5T GE CV/i MRI scanner at $0.3 \times 0.3 \times 0.9$ mm³ voxel resolution. The seeding resolution was chosen to match the voxel resolution for a size of $256 \times 256 \times 128$. Whole-heart fiber tractography was performed using the Runge-Kutta integration scheme (RK4)[31] with a step size of 0.1 mm, a minimum FA threshold of 0.2, and a maximum angle of 45° between consecutive steps. Fiber tracts shorter than 40 voxel units in length were excluded from further analysis resulting in 403,667 fiber tracts.

Cardiac fiber analysis lacks universally accepted anatomical landmarks and readily applicable metrics due to the heart's dynamic microstructural and complex motion. Therefore, we calculated pointwise Helical Angle (HA), quantifying the transmural change in myocyte orientation, and Transmural Depth (TD) representing the normalized distance from the endocardium to the epicardium to provide essential information about location and geometry of fibers within the myocardial wall.

We manually approximate the left ventricle's center to define a radial vector for each streamline point. The local circumferential direction is derived through cross product between radial and z-normal, and local tangent through differences between preceding and succeeding points. The primary diffusion eigenvector is then projected onto the circumferential and tangent directions to compute the helical angle (**Fig. 2**).

TD was derived by solving a Laplace equation with Dirichlet boundary conditions on a tetrahedral mesh of the myocardium, following the approach presented in [3]. The endocardial and epicardial surfaces were labeled 0 and 1 respectively, and the problem was solved using a standard finite element method (FEM) using the DOLFINx library [1, 2, 19, 20]. The solution was mapped back onto each fiber point, yielding a normalized transmural distance from the endocardium.

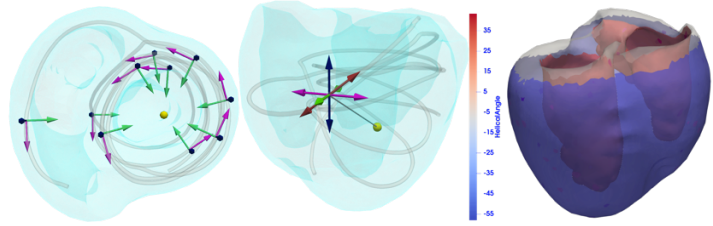

**Fig. 2.** (left) Reference vectors for HA calculations: radial (green), circumferential (pink), z-normal (blue), (middle) tangent vector showing calculated HA; (right) Laplace Dirichlet boundaries used for transmural depth calculation on the volumetric heart mesh. Epicardium (purple) assigned a value of 1 and endocardium (red) assigned a value of 0

### 2.2 Representation Learning with Deep Learning Models

*Input Data.* Each fiber $F$, is defined as a sequence $\{(x_i, y_i, z_i, HA_i, TD_i)\}_{(0 \leq i < m)}$, where m is the number of points in each fiber, with these five features providing both spatial and orientation cues. A 72-8-20 train-validation-test split (∼290k, 32k, and 80k fibers, respectively) was used to train the models.

*Sequential Learning: Bidirectional Long Short-Term Memory (BLSTM).* We aim to learn latent representations that group similar fiber trajectories closely in an embedding space. Thus, we propose a novel pretext task in which a BLSTM predicts the next 25 points (and respective HA and TD) of a fiber, forcing it to capture both short- and long-range structural patterns and learning expressive, pointwise dependencies. This is a critical aspect given the complexity of cardiac fiber trajectories. Furthermore, the bidirectionality of the model addresses the arbitrary direction of fiber integration.

*BLSTM Architecture.* A 4-layer BLSTM with 256 hidden units per layer received the 5-dimensional input sequences, padded to the maximum sequence length within each batch. The final hidden state was projected to predict the subsequent 25 points and was trained for 100 epochs. Alternative Transformer and simpler BLSTM architectures were evaluated, but our proposed BLSTM demonstrated superior convergence speed and lower validation loss.

*Global representation: Transformer-based Autoencoder (TAE).* A Transformer autoencoder effectively captures global relationships by leveraging self-attention across all points in a fiber, learning a comprehensive latent representation that complements the BLSTM embeddings.

*TAE Architecture.* The input sequences were padded to a length of 591 (number of points in longest fiber) with a sentinel value and attention masks for invalid positions.

A linear projection (5 →128 features), sinusoidal positional encodings, and a 4-layer Transformer [23] (8-head self-attention, 512-dimensional feed-forward, 0.1 dropout) formed the encoder, while a symmetrical decoder reconstructed the original fiber. The model was trained for 50 epochs since it showed signs of convergence around epoch 35. This model outperformed a BLSTM-based autoencoder in reconstruction accuracy with fewer parameters, providing a more efficient global fiber representation.

*Training and Implementation.* Both models were trained on an NVIDIA A100 GPU with batch sizes $256, 128$, optimized using the AdamW optimizer with learning rates $0.001, 10^{-4}$, and used a scheduler (ReduceLROnPlateau: factor = 0.1, patience = 3; min_lr: $10^{-6}$ and $10^{-8}$) to improve convergence. Model training used the mean squared error (MSE) between predicted and ground truth feature vectors, averaged across all points and fibers. While the input features $(x, y, z, HA, TD)$ have different scales, the overall MSE was only used as a training signal for learning representations suitable for clustering, which is our primary objective.

### 2.3  Embedding Fusion and Clustering

For each fiber, we extracted a 512-dimensional embedding from the BLSTM by concatenating the final forward and backward hidden states of the top layer, and a 128-dimensional embedding from the TAE encoder. We applied PCA to reduce the dimensions of the resulting concatenated embeddings from 640 to 65 principal components, preserving ~95% of the variance while capturing both local sequential dependencies and global structural context.

Clustering was done using Hierarchical Density-Based Spatial Clustering of Applications with Noise (HDBSCAN) [4] due to its ability to automatically determine cluster numbers, account for variable densities, and classify outliers. We conducted a grid search to optimize the hyperparameters (min_samples:[10,25,50,100,250,500,750], min_cluster_size:[100, 250, 500, 750, 1000, 1500, 2000, 5000]) implemented using the GPU-accelerated RAPIDs cuML [17] on NVIDIA A100 GPU. We assessed cluster quality using quantitative metrics (Silhouette, Davies–Bouldin, Calinski–Harabasz, noise fraction, and number of clusters, Density-Based Clustering Validation (DBCV) suited to non-convex, variable-density structures), and qualitative validation by mapping resulting labels onto 2D t-SNE [10] projections of the embeddings and respective fibers.

## 3  Results

*Fiber Tractography and Feature Extraction Results.* **Fig. 3** illustrates a sample of fiber trajectories color-coded based on HA, ranging from -90° (right-handed helix) to +90° (left-handed helix) and TD representing the normalized distance from the endocardium (0) to the epicardium (1).

*Representation Learning Results.* The BLSTM achieved an MSE of 0.117 on the test set, accurately modelling local sequential fiber geometry (**Fig. 4** (i)). The TAE achieved an MSE of 0.679, capturing global trajectory shape (**Fig. 4** (ii)).

*Clustering Results.* HDBSCAN with optimized parameters (min_samples = 25, min_cluster_size ranging from 750 to 2000), yielded 33 to 62 clusters, providing a flexible tool for cardiac fiber analysis at varying levels of granularity. We show results for min_cluster_size = 2000 in **Fig. 5** and **Fig. 3** right (2D t-SNE plot of the embeddings). Across the range of min_cluster_size values, a consistently low percentage of fibers (21-24%) were classified as noise, indicating the robustness of the clustering.

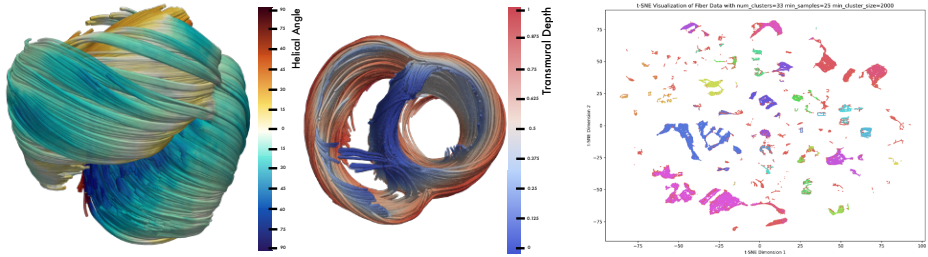

**Fig. 3.** Streamlines color-coded by helical angle (left), and by transmural depth (center). Right: t-SNE plot of learned embeddings

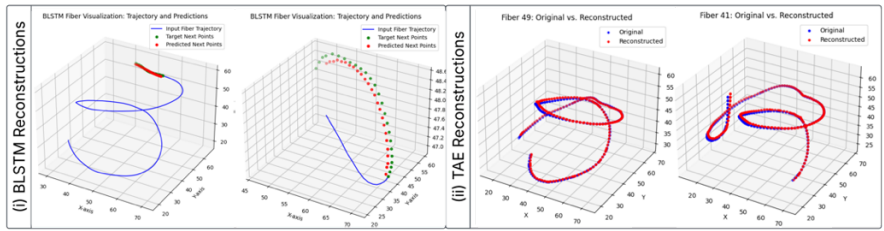

**Fig. 4.** Examples of predicted points by the proposed models

## 4   Discussion

The combined PCA-reduced embeddings exhibited clear structure on the 2D t-SNE plot (**Fig. 3Fig. 5**, right), indicating underlying structures in the fiber populations, while the clustered fibers exhibit distinct spatial arrangements, suggesting that the learned representations successfully captured the subtle changes in fiber orientations despite their spatial proximity and intertwining trajectories.

**Fig. 5** (i) shows two clusters (blue and green) with notable spatial overlap, where the blue bundle almost appears contained within the green from certain angles. If we relied solely on the BLSTM's local sequential cues, these might have merged into a single cluster. However, the TAE leverages global trajectory information and TD, revealing that the green cluster spans a wider transmural range and wraps around more of the left ventricle, indicating a distinct anatomical role compared to the more localized blue cluster.

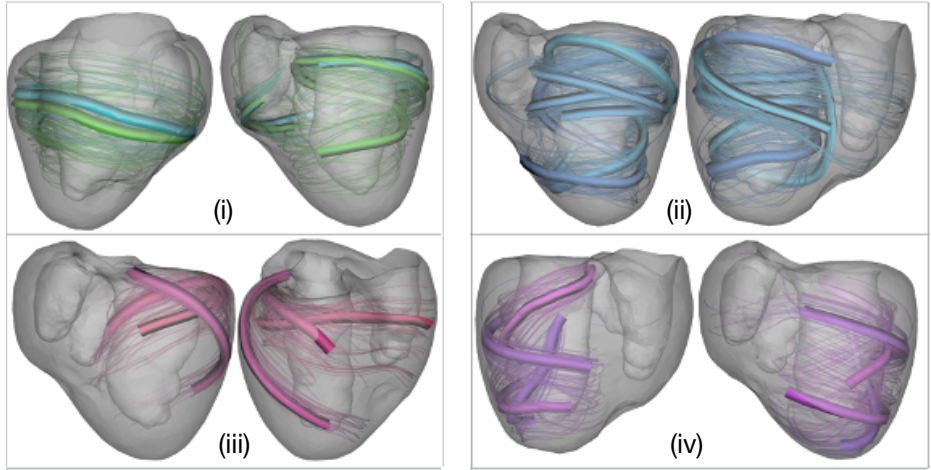

**Fig. 5.** Examples of clustered fibers.

**Fig. 5** (ii) highlights three left ventricular clusters (shades of blue) that share a similar overall shape and have overlapping HA and TD ranges, yet they diverge in their local fiber paths. One spirals uniformly from the apex to the base (darker blue), another remains confined to a narrow z-range (medium blue) and wraps horizontally, and the third (light blue) zigzags along the ventricular wall. While the TAE captures their overall shape similarity, it may struggle to distinguish such fine-grained trajectory details—whereas the BLSTM's sequential learning pinpoints each cluster's unique path. Together, these examples demonstrate how combining local and global representations uncovers subtle differences in fiber organization that would otherwise remain hidden, underscoring the need for both perspectives to accurately characterize myocardial fiber architecture.

**Fig. 5** (iii) shows three pink-shaded clusters near the heart's base that form a braided, intersecting arrangement. Relying on spatial coordinates or overall shape alone would merge them, yet each cluster's HA pattern in the basal region sets it apart. By incorporating HA as a pointwise feature, our method accurately delineates these intertwined trajectories—an achievement that would be difficult with purely spatial or global shape cues. Finally, **Fig. 5** (iv) underscores the culmination of our framework's design by demonstrating how both HA and TD variations separate the epicardial clusters from the endocardial ones, while the BLSTM further refines local differences. The two epicardial clusters share broadly similar HA and TD trends but follow slightly different paths, starting far apart on one side of the heart before wrapping around and nearly intersecting. Conversely, the two endocardial clusters have similar TD profiles yet distinct HA changes that result in clearly diverging trajectories. These findings show that incorporating domain-specific anatomical context improves representation learning, and combining globally varying features (HA, TD) with local sequencing (BLSTM) is crucial for disentangling close yet distinct fiber bundles.

*Limitations and Future Work.* This study focuses on a single, though high-quality, ex-vivo cardiac DTI atlas dataset and further validation on a larger cohort is needed. In addition, DTI tractography is limited in characterizing fiber crossings, such as those seen at the LV/RV junction and apex. Further, the choice of certain parameters in our clustering pipeline involved a degree of subjective judgment, primarily guided by visual inspection. Other datasets may require different parameters.

Our immediate priority is to validate these findings in a multi-patient cohort of human cardiac DTI data. Incorporating secondary and tertiary eigenvectors in our analysis, which was problematic with our dataset [16], would allow us to refine our characterization of laminar structures in the myocardium [18]. Future work should also explore more objective and automated methods for hyperparameter selection in HDBSCAN (e.g., by leveraging its condensed cluster tree or by incorporating anatomical constraints or expert knowledge) to enhance the robustness and generalizability of the method. Finally, we wish to apply this framework to investigate the existence of a global laminar structure in the heart.

## 5     Conclusion

Our work presents a fundamentally new approach to analyzing myocardial fiber architecture from DTI data. By uniquely combining sequential progression information with a global representation and pointwise anatomical context, we have uncovered previously inaccessible details of cardiac architecture. Unlike prior methods limited by localized analyses and averaged metrics, our framework captures the dynamic progression of features like HA and TD along individual fibers, enabling delineation of previously indistinguishable fiber populations. This unprecedented ability to resolve fine-grained cardiac architecture has the potential to advance our understanding of myocyte orientations and their effect on myocardial function and disease.

*Code and Data Availability.* We share our code and data through an anonymized [Google Drive link here](#).

**Disclosure of Interests.** The authors have no competing interests to declare that are relevant to the content of this article.

## References

1. Alnæs, M.S. et al.: Unified form language: A domain-specific language for weak formulations of partial differential equations. ACM Trans Math Softw. 40, 2, 9:1-9:37 (2014). https://doi.org/10.1145/2566630.
2. Baratta, I.A. et al.: DOLFINx: The next generation FEniCS problem solving environment, https://zenodo.org/records/10447666, (2023). https://doi.org/10.5281/zenodo.10447666.
3. Bayer, J.D. et al.: A Novel Rule-Based Algorithm for Assigning Myocardial Fiber Orientation to Computational Heart Models. Ann. Biomed. Eng. 40, 10, 2243–2254 (2012). https://doi.org/10.1007/s10439-012-0593-5.

4. Campello, R.J.G.B. et al.: Density-Based Clustering Based on Hierarchical Density Estimates. In: Pei, J. et al. (eds.) Advances in Knowledge Discovery and Data Mining. pp. 160–172 Springer, Berlin, Heidelberg (2013). https://doi.org/10.1007/978-3-642-37456-2_14.
5. De Luca, A., Leemans, A.: Fiber Tractography Techniques. In: Coriasco, M. et al. (eds.) A Practical Guide to Advanced Diffusion MRI. pp. 111–139 Springer Nature Switzerland, Cham (2024). https://doi.org/10.1007/978-3-031-70337-9_5.
6. Frindel, C. et al.: Cardiac Fibre Trace Clustering for the Interpretation of the Human Heart Architecture. In: Ayache, N. et al. (eds.) Functional Imaging and Modeling of the Heart. pp. 39–48 Springer, Berlin, Heidelberg (2009). https://doi.org/10.1007/978-3-642-01932-6_5.
7. Gupta, V. et al.: FiberNET: An ensemble deep learning framework for clustering white matter fibers, https://www.biorxiv.org/content/10.1101/141036v1, (2017). https://doi.org/10.1101/141036.
8. Lee, S.-E. et al.: Three-dimensional Cardiomyocytes Structure Revealed By Diffusion Tensor Imaging and Its Validation Using a Tissue-Clearing Technique. Sci. Rep. 8, 1, 6640 (2018). https://doi.org/10.1038/s41598-018-24622-6.
9. Legarreta, J.H. et al.: Filtering in tractography using autoencoders (FINTA). Med. Image Anal. 72, 102126 (2021). https://doi.org/10.1016/j.media.2021.102126.
10. Maaten, L. van der, Hinton, G.: Visualizing Data using t-SNE. J. Mach. Learn. Res. 9, 86, 2579–2605 (2008).
11. Mekkaoui, C. et al.: Diffusion MRI Tractography of the Developing Human Fetal Heart. PLOS ONE. 8, 8, e72795 (2013). https://doi.org/10.1371/journal.pone.0072795.
12. Mekkaoui, C. et al.: Diffusion Tractography of the Entire Left Ventricle by Using Free-breathing Accelerated Simultaneous Multisection Imaging. Radiology. 282, 3, 850–856 (2017). https://doi.org/10.1148/radiol.2016152613.
13. Mekkaoui, C. et al.: Myocardial Scar Delineation Using Diffusion Tensor Magnetic Resonance Tractography. J. Am. Heart Assoc. 7, 3, e007834 (2018). https://doi.org/10.1161/JAHA.117.007834.
14. Ngattai Lam, P.D. et al.: TRAFIC: Fiber Tract Classification Using Deep Learning. Proc. SPIE-- Int. Soc. Opt. Eng. 10574, 1057412 (2018). https://doi.org/10.1117/12.2293931.
15. Pashakhanloo, F. et al.: Submillimeter diffusion tensor imaging and late gadolinium enhancement cardiovascular magnetic resonance of chronic myocardial infarction. J. Cardiovasc. Magn. Reson. Off. J. Soc. Cardiovasc. Magn. Reson. 19, 1, 9 (2017). https://doi.org/10.1186/s12968-016-0317-3.
16. Peyrat, J.-M. et al.: A Computational Framework for the Statistical Analysis of Cardiac Diffusion Tensors: Application to a Small Database of Canine Hearts. IEEE Trans. Med. Imaging. 26, 11, 1500–1514 (2007). https://doi.org/10.1109/TMI.2007.907286.
17. Raschka, S. et al.: Machine Learning in Python: Main developments and technology trends in data science, machine learning, and artificial intelligence. ArXiv Prepr. ArXiv200204803. (2020).

18. Rodríguez-Padilla, J. et al.: Impact of intraventricular septal fiber orientation on cardiac electromechanical function. Am. J. Physiol.-Heart Circ. Physiol. 322, 6, H936–H952 (2022). https://doi.org/10.1152/ajpheart.00050.2022.
19. Scroggs, M.W. et al.: Basix: a runtime finite element basis evaluation library. J. Open Source Softw. 7, 73, 3982 (2022). https://doi.org/10.21105/joss.03982.
20. Scroggs, M.W. et al.: Construction of Arbitrary Order Finite Element Degree-of-Freedom Maps on Polygonal and Polyhedral Cell Meshes. ACM Trans Math Softw. 48, 2, 18:1-18:23 (2022). https://doi.org/10.1145/3524456.
21. Sosnovik, D.E. et al.: Diffusion Spectrum MRI Tractography Reveals the Presence of a Complex Network of Residual Myofibers in Infarcted Myocardium. Circ. Cardiovasc. Imaging. 2, 3, 206–212 (2009). https://doi.org/10.1161/CIRCIMAGING.108.815050.
22. Sosnovik, D.E. et al.: Microstructural Impact of Ischemia and Bone Marrow–Derived Cell Therapy Revealed With Diffusion Tensor Magnetic Resonance Imaging Tractography of the Heart In Vivo. Circulation. 129, 17, 1731–1741 (2014). https://doi.org/10.1161/CIRCULATIONAHA.113.005841.
23. Vaswani, A. et al.: Attention Is All You Need, http://arxiv.org/abs/1706.03762, (2023). https://doi.org/10.48550/arXiv.1706.03762.
24. Wang, Z. et al.: Accurate corresponding fiber tract segmentation via FiberGeoMap learner with application to autism. Cereb. Cortex. 33, 13, 8405–8420 (2023). https://doi.org/10.1093/cercor/bhad125.
25. Wilson, A.J. et al.: Myocardial mesostructure and mesofunction. Am. J. Physiol. - Heart Circ. Physiol. 323, 2, H257–H275 (2022). https://doi.org/10.1152/ajpheart.00059.2022.
26. Yongfeng Yuan et al.: Myocardial fiber track-oriented analysis based on the FCM method. 2011 3rd Int. Conf. Adv. Comput. Control. 520–523 (2011). https://doi.org/10.1109/ICACC.2011.6016467.
27. Zhang, F. et al.: Deep white matter analysis (DeepWMA): fast and consistent tractography segmentation. Med. Image Anal. 65, 101761 (2020). https://doi.org/10.1016/j.media.2020.101761.
28. Zhong, S. et al.: Auto-encoded Latent Representations of White Matter Streamlines for Quantitative Distance Analysis. Neuroinformatics. 20, 4, 1105–1120 (2022). https://doi.org/10.1007/s12021-022-09593-4.
29. [2107.04938] Deep Fiber Clustering: Anatomically Informed Unsupervised Deep Learning for Fast and Effective White Matter Parcellation, https://arxiv.org/abs/2107.04938, last accessed 2025/01/25.
30. Helical Structure of the Cardiac Ventricular Anatomy Assessed by Diffusion Tensor Magnetic Resonance Imaging With Multiresolution Tractography. Rev. Esp. Cardiol. Engl. Ed. 66, 10, 782–790 (2013). https://doi.org/10.1016/j.rec.2013.04.021.
31. Press, W.H. et al.: Numerical Recipes 3rd Edition: The Art of Scientific Computing. Cambridge University Press (2007).